# Pressure dependence of magnetic and superconducting transitions in sodium cobalt oxides $Na_xCoO_2$

Y. V. Sushko,[1] O. B. Korneta,[1] S. O. Leontsev,[1] R. Jin,[2] B. C. Sales,[2] and D. Mandrus[2]

[1]*Department of Physics & Astronomy, University of Kentucky, Lexington, Kentucky 40506, USA*

[2]*Condensed Matter Sciences Division, Oak Ridge National Laboratory, Oak Ridge, TN, 37831, USA*

The results of DC magnetization measurements under hydrostatic (helium-gas) pressure are reported for an ambient pressure superconductor $Na_{0.35}CoO_2 \cdot 1.4D_2O$ and its precursor compound, the gamma-phase $Na_{0.75}CoO_2$ that is known to combine a metallic conductivity with an unusual magnetic state below ~22K. The obtained data allowed us to present for the first time the pressure dependence of the magnetic transition in a metallic sodium cobaltate system. This dependence appears to be positive, with the magnetic transition rapidly shifting towards higher temperatures when an applied pressure increases. We ascribe the observed effect to the pressure-induced enhancement of the out-of-plane antiferromagnetic coupling mediated by localized spins interactions (of either superexchange or RKKY type), the scenario consistent with the A-type antiferromagnetic state suggested by recent neutron-scattering data. As for the pressure effect on the superconductivity in $Na_{0.35}CoO_2 \cdot 1.4D_2O$, our measurements established negative and linear for the entire pressure range from 1 bar to 8.3 kbar pressure dependence of $T_c$, the behavior quite different from the reported by previous workers strong non-linearity of the $T_c(P)$ dependence.
(Dated September 12, 2005)

## 1. INTRODUCTION

Due to the recent discovery[1] of superconductivity in $Na_{0.35}CoO_2 \cdot 1.3H_2O$, the whole family of sodium cobalt oxides appeared in a focus of intensive theoretical and experimental studies. The diverse physical properties of the series $Na_xCoO_2$ (x ≤ 1) include, besides superconductivity in hydrated $Na_xCoO_2 \cdot yH_2O$ (x ~ 0.3, y~1.3), also a complex magnetic and transport behavior of the x ~ 3/4 non-hydrated compound in which antiferromagnetic order combines with metallic conductivity[2-5], a "Curie-Weiss metal" state in the x = 2/3 compound, with abnormally large, spin-dependent thermopower[6,7], and an insulating low-temperature state in the x = 1/2 compound with charge ordering on the $CoO_2$ and Na planes[8,9].

Although sodium cobaltates all share the same main structural motive (that is, alternation of layers of edge-sharing $CoO_6$ octahedra with the layers of Na ions), compounds with different Na content crystallize into distinct crystallographic phases, such as the so-called α, α', β, and γ phases that were first described by Fouassier et al. more than 30 years ago[10].

In the first part of this paper we study magnetic properties of floating zone (FZ) grown single crystals of the hexagonal γ-phase, with x ~ 0.75 (space group *P63/mmc*, **a** = 2.83 Å, **c** = 10.85 Å). In this compound, Na atoms occupy two crystallographically non-equivalent positions, Na1 and Na2, with the average relative occupancy of Na1~0.25 and Na2~0.5. Recent studies[4,5] of the gamma-phase single crystals have revealed a broad and hysteretic structural transition that takes place at temperature above the room temperature. In the ordered phase (T < 320 K), the sodium positions Na1 and Na2 were reported to be well defined, as (0, 0, 0.25) and (0.667, 0.333, 0.25), respectively, whereas in the high-temperature phase sodium atoms at the Na2 site were found to randomly occupy three different off-center positions. A simple ionic model prescribes for $Na_{0.75}CoO_2$ a mixed-valent state ($Co^{3.25+}$), and the electronic structure calculations[11] suggest a $t_{2g}$ ground state, with a low-spin configuration for both the $Co^{3+}$ (S = 0) and the $Co^{4+}$ (S=1/2) ions. The second phase transition observed in several properties of the material around 22K signals an onset of magnetic ordering.

For polycrystalline samples, this magnetic transition was thoroughly investigated by Motohashi *et al.*[2] using thermodynamic, transport, and magnetization measurements, and also by muon spin rotation experiments of Sugiyama *et al.*[3]. Magnetic properties of FZ grown single crystals, along with transport, thermodynamic and structural properties, were recently studied by Sales et al[4]. The results of these studies can be summarized as follows. In a wide temperature range above approximately 50 K, the susceptibility of $Na_{0.75}CoO_2$ follows the $\chi = \chi_o + C/(T-\Theta)$ dependence (with $\chi_o$ = 9.49x10$^{-5}$ cm$^3$/mol Co, $p_{eff}$ = 1.56 $\mu_B$, and $\Theta$ = -220 K for *H//***a**, and $\chi_o$ = 2.76x10$^{-4}$ cm$^3$/mol Co, $p_{eff}$ =1 $\mu_B$, and $\Theta$ = -167 K for *H//***c**). For zero field cooling processes, a peak in the $\chi$ vs. *T* dependence is observed at 22 K, suggesting an antiferromagnetic state with the Neel temperature $T_N$ = 22 K. However, a field cooling process in small and moderate fields results in an abrupt increase of the susceptibility below $T_N$, for both in-plane and out-of-plane orientations of magnetic field. Isoterm magnetization curves for both *H*//**a** and *H*//**c** directions exhibit hysteresis typical for ferromagnets, with fairly small value of remanent magnetization (~0.0001$\mu_B$ per Co at T = 1.8 K), suggestive of a weak ferromagnetism. Rather complex nature of the magnetic state of $Na_{0.75}CoO_2$ is further highlighted by the observation that the field cooling process could also result in a peak in susceptibility at $T_N$, provided that a field as high as 5 tesla is applied along the c-axis.

Both superconductivity and magnetism of $Na_xCoO_2$ compounds originate within the $CoO_2$ layers and thus the relationship between the two different types of ordering in this system may have rather intimate character. In attempts to understand what role in the superconductivity of cobalt oxide layers is played by magnetic interactions and magnetic order, the nature and driving forces of the latter must be established. Currently, an itinerant aniferromagnetism of an SDW origin, first suggested for the x~0.75 compound by Motohashi *et al.*[2] and Sugiyama *et al.*[3], appears to be widely accepted. Also, an important new information was recently obtained due to observations[12,13] of incommensurate spin fluctuations in the x = 0.75 and closely related x = 0.82 compounds, the results indicating that FM in-plane and AFM inter-plane interactions are both important in these materials, and suggesting a so-called A-type spin-structure for antiferromagnetic ordering. Nevertheless, in view of the absence of the evidences of a commensurate magnetic structure in neutron scattering data (including recent measurements on a ~1cm$^3$ single crystal[4]) a mechanism responsible for magnetic ordering in metallic sodium cobaltates still remains unknown and thus requires further investigations.

Measuring the pressure dependence of the magnetic transition temperature may provide the direct insight into the ordering mechanism and its relationship to the electronic structure. Indeed, in solids, an external pressure is known to control the values of bond lengths and bond angles as well as degree of overlap between electron orbitals, with a shift and/or split of the energy levels and changes in exchange interaction strength in response on pressure variations. Despite that some pressure/volume dependent measurements on $Na_xCoO_2$ were previously reported[14,15], we are still lacking pressure-dependent data on static susceptibility of $Na_xCoO_2$. In the present work, the pressure dependencies of the 22K magnetic transition in the FZ grown crystals[4] of the γ-phase of $Na_xCoO_2$ with $x\sim0.75$ and the superconducting transition temperature $T_c$ of $Na_{0.35}CoO_2\cdot 1.4D_2O$ polycrystalline sample[16] are obtained by means of the high-resolution (SQUID magnetometer) magnetization measurements under hydrostatic (helium-gas) pressure.

## 2. EXPERIMENTAL

The single crystalline $\gamma$-$Na_{0.75}CoO_2$ and polycrystalline $Na_{0.35}CoO_2\cdot 1.4D_2O$ samples were synthesized in Oak Ridge and their preparation, as well as physical properties at ambient pressure, were in detail reported elsewhere[4,16].

The temperature- and field- dependencies of the magnetization were measured with a commercial MPMS-5 (Quantum Design™) SQUID magnetometer. For measurements under applied pressure we combined the MPMS magnetometer with a helium-gas high-pressure system in a set-up identical to that described in connection to the earlier pressure-dependent studies[17,18] of magnetic susceptibility of the CMR pyrochlore $Tl_2Mn_2O_7$ and double-layer ruthenium oxide perovskite $Sr_3Ru_2O_7$. The main advantage of using helium as a pressure transmitting medium is a truly hydrostatic, and easily controlled during the cooling/warming cycles, pressure. In our experiments, a single crystal sample was placed inside a specially designed long and slim pressure cell (8.6 mm OD, 3.6 mm ID, length 180 mm) connected, via a 3-meter long capillary tubing, to a U11 (Unipress™) gas-compressor system. Both the pressure cell and the capillary were made of BeCu alloy. The pressure cell was placed into the sample chamber of the SQUID magnetometer, and the BeCu capillary tube (3.0 mm OD/0.3 mm ID) not only provided a continuous flow of the pressure medium (helium) into the pressure cell, but also performed a role of the MPMS system's transport rod. Inserted into the pressure cell a 130 mm - long polyimid tube (3 mm in diameter and 40 μm in thickness), with two tiny pieces of a pressed cotton held inside by friction, was used as a sample holder. With such an arrangement, the sample in our experiments is sandwiched between the two soft cotton slabs, in a properly chosen orientation with respect to the direction of the applied magnetic field. Continuous pumping of pressure medium into the pressure cell during the cooling/warming cycles and avoiding any kind of glue for sample mounting prevent shear stresses in the sample and thus ensures that the SQUID picks up a homogeneous sample response.

## 3. PRESSURE EFFECT ON MAGNETIC TRANSITION IN THE GAMMA-PHASE $Na_{0.75}CoO_2$

The temperature dependence of the magnetic moment along **c**-axis of $Na_{0.75}CoO_2$ measured under initial value of pressure of 50 bar is presented in the left panel of Figure 1. The raw data, that include the sample signal plus the negligible small background from the pressure cell, are shown. The behavior in a narrow temperature interval 15-30 K in the vicinity of $T_N$ is only displayed in the figure in order for the details of the magnetic transition to be clearly seen. In agreement with the reported data for ambient pressure[4], we observe that in modest external fields the 22 K transition manifests itself differently for FC and ZFC processes. With the measuring field of 200 Oe, the FC branch of the $m(T)$ dependence is found to exhibit a steep increase below the Neel temperature of 22.1K. As the data in a right panel of Figure 1 demonstrate, a turnover to the high-field regime[4] of $m(T)$, with magnetization decreasing on cooling across $T_N$, happens not in a sudden manner but rather continuously, in a broad field domain below 2.8 tesla. Another interesting effect of high magnetic fields is a "shoulder" seen in the $m$ vs. $T$ dependence at the temperature near 10 K. In somewhat lower fields, i.e. at 2 T, a precursor for this feature is seen in a form of a rather rounded maximum at the same temperature of ~10 K. The origin of the 10 K anomaly in the $m(T)$ dependence is currently not understood and thus requires further studies. We speculate, that it may reflect a low-temperature transformation of the magnetic structure with spin reorientation. Detailed measurements of the field- and temperature- dependencies of magnetization along different crystallographic directions in *precisely oriented* magnetic fields are expected to shed light on such a possibility.

The evolution with pressure of the temperature dependence of DC magnetization of the $Na_{0.75}CoO_2$ single crystal is shown in Fig. 2. The main panel presents the $m$ vs. $T$ data obtained at four different pressures during the field cooling processes with the field H =200 Oe applied along the **c** axis. The main effect of pressure is observed in a shift of the transition curve to the higher temperatures, with increase of external pressure from 50 bar to 8.65 kbar resulting in a shift of the onset point of transition from ~ 22.1 K to ~ 24. 3 K — an appreciable 10% growth of $T_N$. The values of $T_N$ that we recorded at seven different pressures and plotted against $P$ in the inset figure allows to summarize that in overall, the magnetic transition temperature of $Na_{0.75}CoO_2$ follows linear pressure dependence with a fairly large baric coefficient $dT_N/dP = 0.25$ K/kbar.

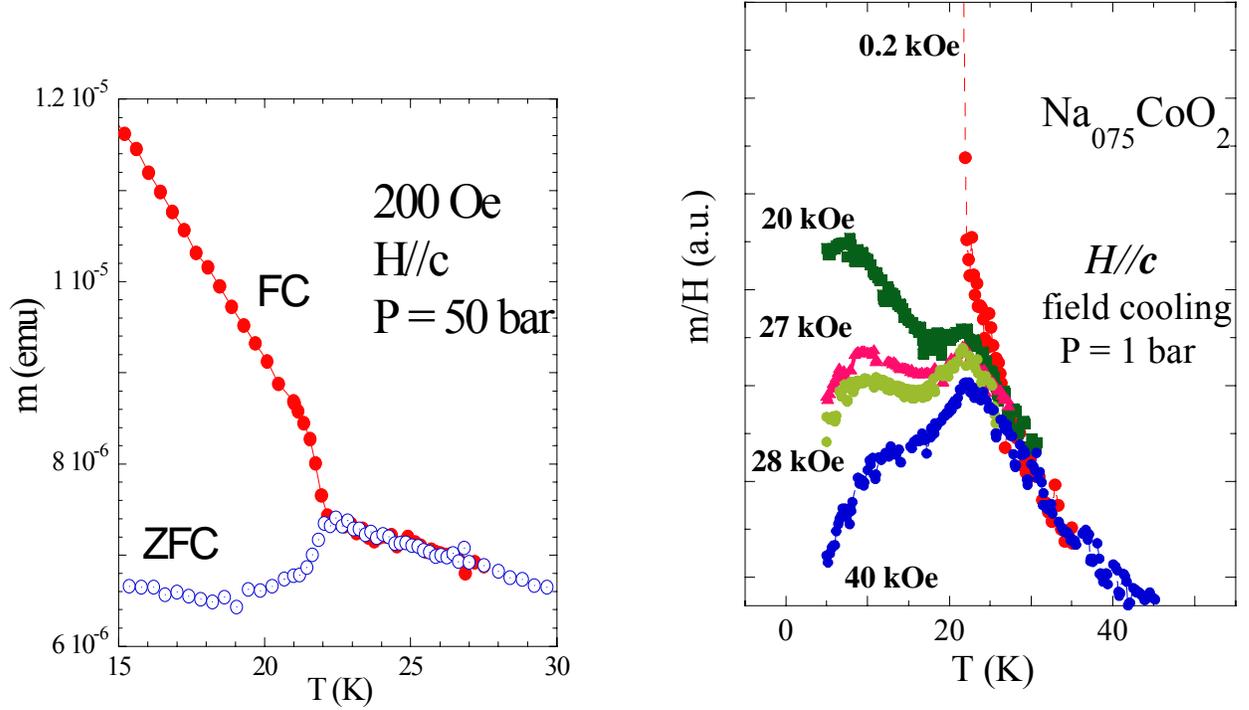

FIG.1. (Color online). Left: ZFC and FC branches of the temperature dependence of the dc magnetic moment of a single crystal of $Na_{0.75}CoO_2$ measured under a slightly elevated external pressure of 50 bar with the field of 200 Oe applied perpendicular to the *ab* plane. The data are shown without correction for a background signal of the pressure cell. Right: the temperature dependence of the FC static magnetic susceptibility $\chi = M/H$ for three different values of an external magnetic field applied perpendicular to the *ab* plane of the crystal.

The observed positive pressure effect on magnetic transition is definitely the one that was least expected for the $Na_{0.75}CoO_2$ in view of rather wide acceptance among current workers of an SDW mechanism of magnetic ordering in this system. Indeed, it is well known (chromium and organic Bechgaard salts $TMTSF_2X$ are among the textbook examples) that application of pressure causes detrimental effect on SDW state. Interpretation of this fact is rather straightforward – the SDW formation requires the presence of congruent (nested) parts on the Fermi surface (FS) and while perfect nesting is a property of electronic band structure of one-dimensional (1-D) or quasi-one-dimensional (Q-1-D) systems with cylindrical FS, the nesting property should deteriorate due to the increased transverse dispersion in cases when the dimensionality of a system increases, going from 1-D to 2-D and 3-D. Since an external pressure is naturally expected to suppress the low-dimensionality effects, it should also suppress the FS nesting and thus density-wave state.

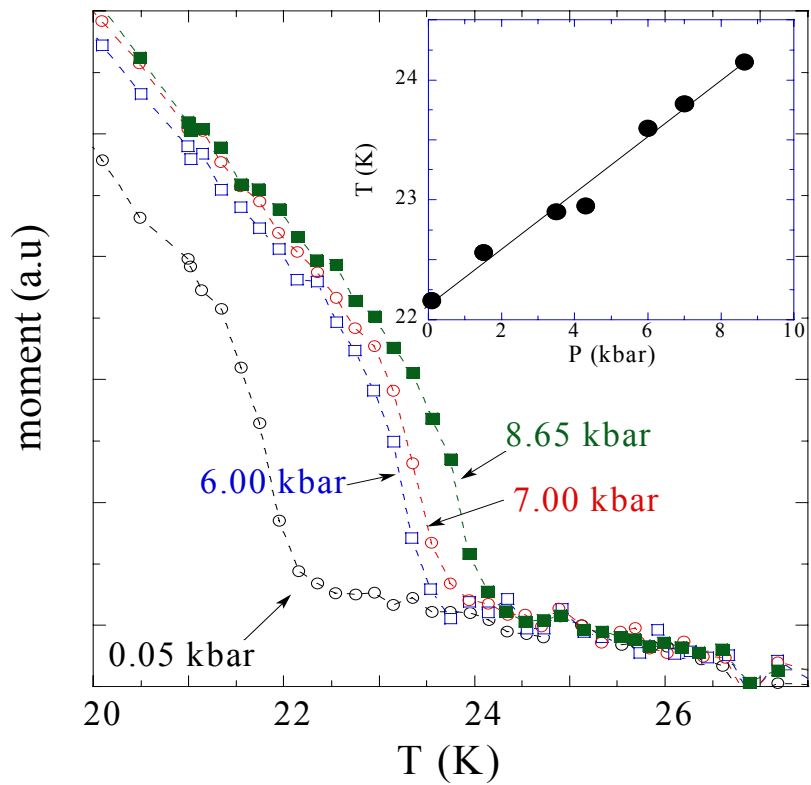

FIG.2. (Color online). Field cooled magnetization of the FZ grown crystal of $Na_{0.75}CoO_2$ (measured in a magnetic field of 200 Oe with H//c) as a function of temperature and pressure. Inset: the pressure dependence of the magnetic transition temperature $T_N$. The lines are guides for the eye.

The observed in our experiments positive pressure shift of $T_N$ thus appears to deny a purely band (SDW) model of antiferromagnetism. As an alternative interpretation, we suggest a combination of itinerant and localized moment mechanisms, a scenario in which strongly anisotropic, layered character of $Na_{0.75}CoO_2$, with a rather suppressed interlayer transport (as indicated by the very large, as high as 800, anisotropy ratio $\rho_c/\rho_{ab}$ [4]) plays a crucial role. Namely, an A-type antiferromagnetism with the ferromagnetic correlation of the itinerant spins within the highly-conducting $CoO_2$ planes and either superexchange- or Ruderman-Kittel-Kasuya-Yosida (RKKY) interaction- mediated antiferromagnetic inter-plane coupling could account well for the observed positive pressure effect. If the localized spins are indeed present in $Na_{0.75}CoO_2$ (we remind here that susceptibility follows a Curie–Weiss law in a fairly wide temperature range[2,4]), they may play significant role in magnetic interactions via the interlayer superexchange between Co moments. According to a recent theoretical work of Johannes et al.[19], several indirect, sodium-containing Co-O-Na-O-Co bonds appear to be the most effective ones among possible interlayer superexchange paths within the sodium cobalt oxide structure. We expect these bonds to be very sensitive to the crystal volume changes and thus attribute the increase of $T_N$ under hydrostatic pressure to the increased strength of the interlayer coupling due to compression of the superexchange bondlength. Furthermore, since the RKKY exchange term $J(\mathbf{r})$ is distance dependent, the lattice compression could lead to $T_N$ increasing with pressure also when the RKKY interaction, and not superexchange, predominantly affect the regime of the interlayer spin-coupling.

## 4. PRESSURE DEPENDENCE OF SUPERCONDUCTING TRANSITION TEMPERATURE IN $Na_{0.35}CoO_2 \cdot 1.4D_2O$

The temperature dependence of the static susceptibility of superconducting compound $Na_{0.35}CoO_2 \cdot 1.4D_2O$ at different pressures is depicted in Figure 3. The main panel shows a plot of susceptibility versus temperature under pressures of 0.33 to 8.30 kbar in the temperature range 1.7-15 K. The data reveal that a 25-fold increase in external pressure has no effect on the magnitude of both the ZFC (shielding) and the FC (Meisner effect) diamagnetic signals. This result is worth mentioning because it evidences that a superconducting volume fraction of the hydrated (deuterated) layered sodium cobalt oxide, although reportedly rather undeveloped at ambient pressure (the work of Takada and coworkers[1] gave an estimate of the superconducting fraction as ~ 6.5% of the corresponding to perfect diamagnetism value) does not grow under pressure. Rather strong effect of applied pressure is however observed for the superconducting transition temperature. The inset to the Figure 3, in which we plotted, as a function of temperature, the normalized values of susceptibility for four different pressures, clearly shows that the superconducting transition temperature decreases with increasing pressure. The overall $T_c$ vs. $P$ dependence, as derived from the data for the onset temperature of the transition at fourteen different values of applied pressure between 1 bar and 8.3 kbar, is presented in Figure 4.

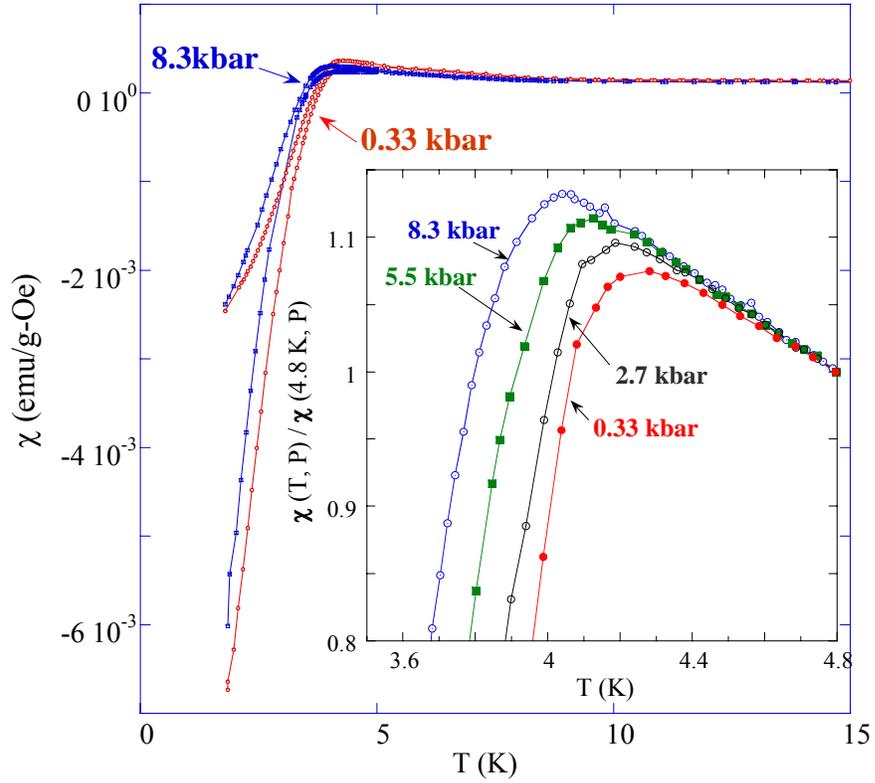

FIG. 3. (Color online). The ZFC and FC static magnetic susceptibility versus temperature under pressures of 0.3 and 8.3 kbar. Inset: the details of temperature dependence of susceptibility in the vicinity of the superconducting transition onset point for four different pressures.

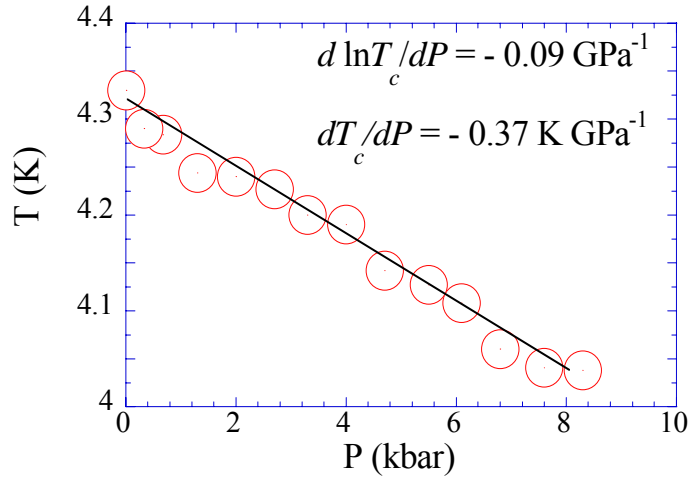

FIG. 4. Pressure dependence of superconducting critical temperature $T_c$ of a polycristalline $Na_{0.35}CoO_2 \cdot 1.4D_2O$.

It is instructive to compare our results with the $T_c$ (P) data of Lorenz et al.[14], that were obtained on a polycrystalline sample of $Na_{0.3}CoO_2 \cdot 1.3H_2O$ by means of AC susceptibility measurements with the clamp cell technique. Although the magnitudes of the overall pressure shift of the superconducting critical temperature appear to be rather close (compare our value of -0.09 $GPa^{-1}$ with -0.07 $GPa^{-1}$ reported by Lorenz[14]), and the sign is negative, an important difference is noticeable. Namely, our data reveal that the $T_c$ varies linearly with P, whereas the Ref. 14 emphasizes strong nonlinearity in the $T_c$ vs. P relation. Such a difference may reflect different sample quality of the polycrystalline samples used in both studies, but also could be caused by substantial deviations from hydrostatic pressure conditions in the clamped cell experiments due to non-monotonic variation of pressure in the cooling process, the effect, as shown by previous workers[20,21], especially strong in an initial, a few kilobar range of pressures.

We also would like to comment that negative pressure dependence of $T_c$ should be naïvely expected for sodium cobalt oxide system where the very occurrence of superconductivity requires large interlayer separation (the result of intercalation), and also agrees well with the simple BCS picture of the phonon-mediated superconductivity (where gradual suppression of $T_c$ is attributed to the decrease in electron-phonon coupling strength, due to stiffening of the lattice and/or decrease of the electron density of states). Concerning, however, possible connection between magnetism and superconductivity in sodium cobaltates, we note that the observed under applied pressure suppression of superconductivity in $Na_{0.35}CoO_2 \cdot 1.4D_2O$ together with enhancement of the magnetism in the parent compound $Na_{0.75}CoO$ could be interpreted as indication that in layered cobalt oxides magnetic interactions play a traditional (negative) effect on superconductivity.

## 5. SUMMARY AND CONCLUSIONS

In conclusion, we reported the results of the first study of the effect of external pressure on the magnetic transition in the metallic gamma-phase compound $Na_{0.75}CoO_2$. We also investigated in detail the behavior under pressure of the polycrystalline samples of superconducting material $Na_{0.35}CoO_2 \cdot 1.4D_2O$, particularly in the low pressure region for which previously reported ac susceptibility measurements under quasi-hydrostatic pressure conditions (clamp cell method ) revealed the non-linear pressure dependence of the $T_c$. Concerning superconductivity in hydrated/deuterated sodium cobalt oxides, the observed in the present work linear pressure dependence of the $T_c$ appears to be an important new result . For the $Na_{0.75}CoO_2$ single crystals, our magnetization measurements revealed a strong and linear with pressure increase of transition temperature $T_N$ in a wide range of pressures up to 8.65 kbar, the highest pressure in our experiments. To account for this result, we propose a simple model in which the antiferromagnetic ordering in γ-$Na_{0.75}CoO_2$ is viewed as an A-type AFM state, with the ferromagnetic correlations of the itinerant spins within the $CoO_2$ planes, and localized spins mediated (via either superexchange or RKKY interaction), and thus strongly distance-dependent, antiferromagnetic coupling between the planes.

## ACKNOWLEDGEMENTS

Work at University of Kentucky was supported in part by NSF (Grant No. DMR 05-02706). Oak Ridge National Laboratory is managed by UT-Battelle, LLC, for the U.S. Department of Energy under contract DE-AC05-00OR22725.

---

1. K. Takada, H. Sakural, E. Takayama-Muromachi, F. Izumi, R.A. Dilanlan, and T. Sasaki, *Nature(London)* **422**, 53 (2003).

2. T. Motohashi, R. Ueda, E. Naujalis, T. Tojo, I. Terasaki, T. Atake, M. Karppinen, and H. Yamauchi, *Phys. Rev. B* **67**, 064406 (2003).

3. J. Sugiyama, H. Itahara, J. H. Brewer, E. Ansaldo, T. Motohashi, M. Karppinen, and H. Yamauchi, *Phys. Rev. B* **67**, 214420 (2003).

4. B. C. Sales, R. Jin, K. A. Affholter, P. Khalifah, G. M. Veith, D. Mandrus, *Phys. Rev. B* **70**, 174419 (2004).

5. Q. Huang, J. W. Lynn, B.H. Toby, et al., *J. Phys.-Cond.Mat.* **17**, 1831 (2005).

6. I. Terasaki, Y. Sasago, and K. Uchinokura, *Phys. Rev. B* **56**, 12 685 (1997).

7. Y. Wang, N. S. Rogado, R. J. Cava, and N. P. Ong, *Nature (London)* **423**, 425 (2003).

8. M. L. Foo, Y. Wang, S. Watauchi, H. W. Zandbergen, T. He, R. J. Cava, and N. P. Ong , *Phys. Rev. Lett.* **92**, 247001 (2004).

9. Y. J. Uemura, P. L. Russo, A.T. Savici, C. R. Wiebe, G. J. MacDougall, G. M. Luke, M. Mochizuki, Y. Yanase, M. Ogata, M. L. Foo, R. J. Cava, *cond-mat/0403031*.

10. C. Fouassier, G. Matejka, Jean-Maurice Reauet, P. Hagenmuller, *J. Solid State Chem.* **6**, 532 (1973).
11. D.J. Singh, *Phys. Rev. B* **68**, 020503 (2003).
12. L. M. Helme, A. T. Boothroyd, R. Coldea, D. Prabhakaran, D. A. Tennant, A. Hiess, and J. Kulda, *Phys. Rev. Lett.* **94**, 157206 (2005).
13. S. P. Bayrakci, I. Mirebeau, P. Bourges, Y. Sidis, M. Enderle, J. Mesot, D. P. Chen, C. T. Lin, and B. Keimer, *Phys. Rev. Lett.* **94**, 157205 (2005).
14. B. Lorenz, J. Cmaidalka, R. L. Meng, and C. W. Chu**,** *Phys. Rev. B* **68**, 132504 (2003).
15. S. Park, Y. Lee, A. Moodenbaugh, and T. Vogt, *Phys. Rev. B* **68**, 180505 (2003).
16. R. Jin, B. C. Sales, P. Khalifah, and D. Mandrus, *Phys. Rev. Lett.* **91**, 217001 (2003).
17. Yu. V. Sushko, Y. Kubo, Y. Shimakawa, and T. Manako, *Rev. High Pressure Sci. Technol,* **7,** 505 (1998); *Physica B* **259-261**, 831 (1999).
18. Y. V. Sushko, B. DeHarak, G. Cao, G. Shaw, D. K. Powell, and J. W. Brill, *Solid State Commun.* **130**, 341 (2004).
19. M. D. Johanes, I. I. Mazin, and D. J. Singh, *Phys. Rev. B* **71**, 214410 (2005).
20. N. B. Brandt, S. V. Kuvshinnikov, N. Ya. Minina and E. P. Skipetrov, *Cryogenics* **14**, 464 (1974).
21. J. D. Thompson, *Rev. Sci. Instrum.* **55**, 231 (1983).